\documentstyle[twocolumn,prb,aps,epsf]{revtex}

\begin{document}
\twocolumn[
\title{Bcc $^4$He as a Coherent Quantum Solid}
\author{Nir Gov}
\address{Physics Department,\\
Technion-Israel Institute of Technology,\\
Haifa 32000, Israel}
\maketitle
\tightenlines
\widetext
\advance\leftskip by 57pt
\advance\rightskip by 57pt

\begin{abstract} 
In this work we investigate implications of the quantum nature of bcc $^{4}$%
He. We show that it is a unique solid phase with both a lattice structure 
and an Off-Diagonal Long Range Order of coherently oscillating local 
electric dipole moments. These dipoles arise from the local motion of the 
atoms in the crystal potential well, and oscillate in synchrony to reduce 
the dipolar interaction energy. The dipolar ground-state is therefore found 
to be a coherent state with a well defined global phase and a 
three-component complex order parameter. The condensation energy of the 
dipoles in the bcc phase stabilizes it over the hcp phase at finite 
temperatures. We further show that there can be fermionic excitations of 
this ground-state and predict that they form an optical-like branch in the 
(110) direction. A comparison with 'super-solid' models is also discussed. 
\end{abstract} 

\vskip 0.2cm
]

\narrowtext
\tightenlines

\vspace{.2cm}

\section{Introduction} 
 
The bcc phase of $^{4}$He has a pronounced quantum nature due to the 
relatively open structure of the lattice. Quantum effects are manifested in 
strong anharmonicity of some phonon modes and in the large zero-point 
kinetic energy of the atoms \cite{glyde}. It is this large kinetic energy 
which is thought to help stabilize the bcc phase over the hcp phase. In this 
paper we highlight this nature of the bcc phase by proposing a new physical 
model for the local atomic motion. For the sake of clarity we have 
reproduced here some of the arguments and calculations already given in \cite 
{niremil}. We propose that in the bcc $^{4}$He phase the local excitations 
of the atoms in their potential wells, result in oscillating local electric 
dipoles. The ground-state of these dipoles has the dipoles oscillating in 
synchrony, thereby reducing the dipolar interaction energy between them. 
Solving a mean-field Hamiltonian describing these dipoles we find that 
Bosonic phase fluctuations in the (110) direction reproduce the spectrum of 
the T$_{1}(110)$ phonon. 
 
In the following we further explore the nature of the coherent ground-state 
of the local-modes in the bcc $^{4}$He. We show that the bcc $^{4}$He is a 
unique phase having both Diagonal Long Range Order (DLRO) of the solid 
lattice and Off-Diagonal Long Range Order (ODLRO) of the local dipoles. 
There is therefore a complex three-component order parameter which describes 
the coherently oscillating dipoles in each of the three orthogonal 
directions in the lattice. In the ground-state the local dipoles form a 
Bose-Einstein condensate in the zero momentum state, and we are able to 
estimate the ground-state energy reduction due to this condensation. This 
estimate compares favourably with experimental results and consequently we 
claim that this condensation energy stabilizes the bcc phase over the hcp 
phase. In the hcp phase we expect no coherence or condensation due to the 
highly isotropic lattice and the geometric frustration of the hexagonal symmetry. We also comment about the 
relation of this work to previous work about the 'super-solid' concept in 
quantum solids. 
 
Additionally, we predict a high-energy optical-like mode which has fermionic 
statistics. This excitation is confined to the (110) direction and involves 
a local 'fliping' of a dipole with respect to the ground-state. This makes 
the dipole become anti-symmetric (a $\pi $ phase difference) with respect to 
the global phase of the complex order parameter and aquire Fermi-Dirac 
statistics. We also give analytic expressions for the scattering intensity 
of both the Bose and Fermi excitations along the (110) direction. These 
predictions remain to be compared with future experimental data. 
 
\section{Ground-state coherence and Bose excitations.} 
 
The usual treatment of the ground-state and energy of bcc $^{4}$He employs 
variational wavefunctions that aim to account for the short-range 
correlations between the atoms \cite{glyde}. These correlations arise mainly 
due to the hard-core repulsion between the atoms. The atoms have a high 
zero-point kinetic energy which is given quite accurately by treating them 
as independent particles held in place by the potential of the neighboring 
(static) atoms. This type of calculation is the ''particle-in-cell'' 
approximation which gives surprisingly good agreement with measured 
thermodynamic properties of the solid phase \cite{glyde1}. We want to focus 
here on the effects of the local motion of the atoms inside this 
potential-well on the nature of the ground-state. In this approach we would 
like to isolate the lowest energy excited state of the atom inside its 
potential well, and treat it as a local excitation of the lattice. This 
local excited state consists of a local oscillatory motion of the atom along 
a particular direction and produces an oscillating electric dipole, 
similarly to that of the usual Van-der Waals interaction. However unlike the 
case of the Van-der Waals interaction, in which the dipolar fluctuations are 
random, we show that in the bcc solid there are local dipoles which are 
correlated and a new ground-state of lower energy is created. 
 
The potential well of an atom in the bcc lattice due to the standard helium 
pair-potential $\upsilon (r)$ \cite{glyde}, provided one can take the other 
atoms as stationary, can be maped along any direction in the lattice. We 
find \cite{niremil} that in the directions normal to the unit cube's faces 
(i.e. (100),(010) etc.) the confining potential well is very wide with a 
pronounced double-minimum structure (Fig.1). Solving the one-dimensional 
Schrodinger's equation for a $^{4}$He atom in this potential, we get a first 
excited level with energy $10{\rm K}$, and a wavefunction describing atomic 
motion with an amplitude of $\sim 1{\rm \AA }$ (in the (100) direction 
(Fig.1)). Based on the above calculation, we shall assume that the atoms 
have a local-mode that is highly directional along one of the directions 
equivalent to (100). Local atomic motion along the other directions is 
assumed to be severly restricted due to the higher excitation energies 
(Fig.1). Experimental evidence for the existence of such a ''local mode'' 
comes from NMR measurements which find an activation energy of 7$\pm 1$K\cite 
{schuster,allen}, and we propose to identify this local-mode with 
the highly directional motion of the atoms in the normal directions.
 
Using this identification we can now estimate the size of the local electric 
dipole moment that can be created by this local and highly directional 
atomic motion. As the atom moves this instantaneous local electric-dipole is 
created due to the electronic cloud and the ion being slightly displaced 
relative to each other. The electric dipole moment due to mixing of the 
lowest $\left| s\right\rangle $ and $\left| p\right\rangle $ 
electronic-levels of the $^{4}$He atom, is given from perturbation theory as  
\begin{eqnarray} 
\psi &=&\left| s\right\rangle +\lambda \left| p\right\rangle \Rightarrow 
E_{0}\simeq \left\langle \psi \left| E\right| \psi \right\rangle 
-\left\langle s\left| E\right| s\right\rangle \simeq \lambda 
^{2}\left\langle p\left| E\right| p\right\rangle  \nonumber \\ 
\ &\Rightarrow &\lambda ^{2}\simeq 7/2.46\cdot 10^{4}\simeq 0.00284,\lambda 
\simeq 0.0168  \label{lamda} 
\end{eqnarray} 
 
where $\left| s\right\rangle $ and $\left| p\right\rangle $ stand for the 
ground-state and first excited-state of the $^{4}$He atom, $\lambda $ is the 
mixing coefficient and $\left\langle p\left| E\right| p\right\rangle \simeq 
2.46\cdot 10^{4}$ K is the excitation energy of the first atomic 
excited-state \cite{white}. This small estimated mixing gives the magnitude 
of the induced dipole moment as  
\begin{equation} 
\left| {\bf \mu }\right| =e\left\langle \psi \left| x\right| \psi 
\right\rangle \simeq 2e\lambda \left\langle s\left| x\right| p\right\rangle 
\simeq e\cdot 0.03{\rm \AA }  \label{mu} 
\end{equation} 
 
where $\left\langle s\left| x\right| p\right\rangle \simeq 0.9{\rm \AA }$ . 
The estimation of the mixing $\lambda $ and the dipole-moment $\left| {\bf %
\mu }\right| $ serves to set an upper bound on the magnitude of this effect, 
since we assumed that the entire excitation energy $E_{0}$ is converted to a 
local electric dipole. 

It is possible to show that the lowest energy of a correlated dipolar
array
in the bcc lattice preserves the symmetry of the bcc unit cell along one
of
the symmetry axes. In such a case it can be easily shown that there will
be
no contribution to the dipolar interaction energy from oscillating dipole moments
which
are orthogonal, and 
the instantaneous dipolar 
interaction energy for each of the three orthogonal directions, is given by  
\begin{equation} 
E_{dipole}=-\left| {\bf \mu }\right| ^{2}\sum_{i\neq 0}\left[ \frac{3\cos 
^{2}\left( {\bf \mu }\cdot \left( {\bf r}_{0}-{\bf r}_{i}\right) \right) -1}{%
\left| {\bf r}_{0}-{\bf r}_{i}\right| ^{3}}\right]  \label{edipole} 
\end{equation} 
 
where the sum is over all the atoms in the lattice, ${\bf r}_{i}$ being the 
instantaneous coordinate of the $i$-th atom. For oscillating dipoles with 
random phases, the average instantaneous interaction energy (\ref{edipole}) 
summed over the lattice would be zero. However, the energy of the dipoles 
can be lowered by correlating the phases of the oscillating atoms. Since the 
direction of the local dipole shows the instantaneous direction of the 
motion or displacement, a state where all the dipoles point in the same 
direction is just a uniform motion or translation of the entire lattice. We 
therefore have to look for symmetric arrangements with respect to the number 
of up/down dipoles, such as is shown in Fig.2. This is the lowest energy 
'antiferroelectric' configuration with the periodicity of the bcc unit cell. 
We have shown this arrangement for individual dipoles oriented along the 
(001) direction, but they are similarly arranged for dipoles along the two 
other orthogonal axes. The sum in (\ref{edipole}) for such a configuration 
with a unit dipole is given in Fig.2. Thus, the ground state in our picture 
has the atoms executing this correlated local oscillation along the three 
orthogonal directions. 
 
We therefore have, in addition to the usual (isotropic) Van-der Waals interaction, 
highly directional (anisotropic) electric dipoles that become correlated so that they oscillate in synchrony. 
This is a state of quantum resonance where the system oscillates between two 
equivalent up/down arrangements of the ground-state of the dipoles (Fig.2). 
The total interaction between the atoms is now given as the usual 
second-order ($\propto 1/r^{6}$) Van-der Waals contribution that is the 
result of local-dipoles which have random relative phases, and an additional 
long-range (first-order, $\propto 1/r^{3}$) dipolar interaction from the correlated part. 
Dipolar interactions that decay as $1/r^{3}$ occur for perfectly correlated 
oscillating dipoles, such as a single electric dipole and its image in an 
adjacent conducting plate. The coherently oscillating nearest-neighbor 
dipoles therefore behave as perfect images of each other (Fig.2), and 
oscillate with the same global phase. 
 
The correlated oscillating dipoles do not have an average static dipole 
moment, so this is not the case of an antiferroelectric structural phase 
transition \cite{landau}. The array shown in Fig.2 is simultaneously 
arranged along the other two orthogonal axes. Along each direction the 
ground-state is given as a coherent-state of these local dipoles, i.e. has a 
well-defined phase and an ill-defined occupation number. 
 
We shall treat the dynamics of the correlated dipolar array as independent 
of the other degrees of freedom of the lattice. This assumption needs 
justification since there can be phonon modes that will modulate the atomic 
motion, thereby coupling with the oscillating dipolar array. The oscillatory 
atomic motion induced by the phonons will modulate the relative phases of 
the dipoles. Let us look at the ground state of the dipoles, taking for
example dipolar
oscillations oriented along the (001) direction (Fig.2). We now need to 
consider only phonons which will modulate the local motion responsible for 
the oscillating dipoles in this direction. In the bcc structure, only 3 
phonons fulfill this condition: L(001), T(100) and T$_{1}$(110). Let us 
calculate the energy of the dipolar array when modulated along these 3 
directions. For a modulation along some direction ${\bf k}$ , the dipolar 
interaction energy is given by\cite{heller}:  
\begin{eqnarray}
X\left( {\bf k}\right) &=&%
-\left| {\bf \mu }\right| ^{2}\sum_{i\neq 0}\left[%
\frac{3\cos ^{2}\left( {\bf \mu }\cdot \left( {\bf r}_{0}-{\bf
r}_{i}\right)%
\right) -1}{\left| {\bf r}_{0}-{\bf r}_{i}\right| ^{3}}\right] \nonumber
\\
&&\ \ \exp \left[
2\pi i{\bf k}\cdot \left( {\bf r}_{0}-{\bf r}_{i}\right) \right]
\label{xk}
\end{eqnarray}
 
At $k=0$ the interaction matrix $X(k)$ is just the dipolar energy (\ref 
{edipole}). 
 
In Fig.3 we plot the value of $X(k)$, the energy of the dipolar array 
modulated by the relevant phonons:\ L(001), T(100), and T$_{1}$(110), for 
dipole moment $\left| {\bf \mu }\right| =1$. We see that for a modulation by 
L(001) and T(100) the periodicity of $X(k)$ is over a full unit-cell, that 
is twice the periodicity of these phonons. Since symmetric functions of 
periodicities $\pi /a$ and $2\pi /a$ are orthogonal, so are the 
eigenfunctions of these particular phonons and dipole-excitations. The 
dipole array cannot therefore be excited by any of these two phonon. For the 
modulation produced by the T$_{1}$(110) mode, the periodicity of $X(k)$ is 
the same as that of the T$_{1}$(110) phonon, which can therefore couple to 
the dipole array. We conlude therefore that the coupling of the local modes 
to the lattice excitations is limited to a single phonon mode, justifying 
our assumption that the local modes can be treated separately to a good 
approximation. We shall now calculate the dispersion relation of such an 
excitation by a mean-field solution of an effective Hamiltonian. It turns 
out that the only phonon mode of the bcc lattice that can couple with the 
dipolar array is in fact the natural excitation of the dipolar array in the 
(110) direction. Thus, the only elementary (Bose) excitations of the dipole 
array would be in the (110) direction, in the form of the T$_{1}$(110) 
phonon. The description of this phonon is therefore taken into account by 
our treatment of the dynamics of the dipolar array, and will appear as a 
solution of the mean-field treatment. This means that our assumption of an 
effective decoupling between the dipolar and other degrees of freedom is 
justified. 
 
The Hamiltonian treatment of interacting local excitations was developed 
originally by Hopfield \cite{hopfield} for the problem of excitons in a 
dielectric material. The local excitations are treated as bosons using the 
standard Holstein-Primakof procedure, and the effective Hamiltonian 
describing their behavior is \cite{anderson} 
 
\begin{eqnarray}
{H_{loc}} &=&{\sum_{k}}(E_{0}+X(k))\left( {{b_{k}}^{\dagger
}}{b_{k}}+{\frac{%
1}{2}}\right)  \nonumber \\
&&\ \ +{\sum_{k}}X(k)\left( {{b_{k}}^{\dagger }}{b_{-k}^{\dagger }}%
+b_{k}b_{-k}\right)  \label{hloc}
\end{eqnarray}
where ${{b_{k}}^{\dagger },}{b_{k}}$ are Bose creation/anihilation operators 
of the local mode, $X(k)$ is given in (\ref{xk}) and $E_0$ is the energy
of exciting a local dipole out of the correlated ground-state. 
 
The Hamiltonian ${H_{loc}}$ (\ref{hloc}) which describes the effective 
interaction between localized modes can be diagonalized using the Bogoliubov 
transformation ${\beta _{k}}=u(k)b_{k}+v(k)b{^{\dagger }}_{-k}$. The two 
functions $u(k)$ and $v(k)$ are given by:  
\begin{equation} 
{u^{2}}(k)={\frac{1}{2}}\left( \frac{E_{0}{+X(k)}}{{E(k)}}+1\right) ,{v^{2}}%
(k)={\frac{1}{2}}\left( \frac{E_{0}{+X(k)}}{{E(k)}}-1\right)  \label{uv} 
\end{equation} 
 
The result of solving by mean-field the effective Hamiltonian for the 
correlated dipolar array \cite{niremil}, is a coherent ground-state given by  
\cite{huang}  
\begin{equation} 
\left| \Psi _{0}\right\rangle =\prod_{k}\exp \left( \frac{v_{k}}{u_{k}}{{%
b_{k}}^{\dagger }}{b_{-k}^{\dagger }}\right) \left| vac\right\rangle 
\label{psi0} 
\end{equation} 
 
and the energy spectrum is  
\begin{equation} 
E(k)=\sqrt{E_0\left( E_0+2X(k)\right) }  \label{ek} 
\end{equation} 
 
In Fig.2 we see that the ground-state arrangement has the dipoles arranged 
in alternating planes in the (110) direction. As we have shown the only 
naturally occuring Bose excitations of this dipolar field are along this 
direction and $X(k)$ is the dipolar interaction matrix element for $k$ in 
the (110) direction (Fig.3). In order to calculate the energy spectrum we 
now need to fix the size of the coherent dipole moment $\left| {\bf \mu }%
\right| $. According to our definition of the local mode the energy cost of 
flipping the direction of a single dipole out of the ordered ground state is 
defined to be $E_{0}$. This is equivalent to having $2\left| X(k=0)\right| 
=E_{0}$, which is the condition to have a gapless mode at $k\rightarrow 0$ (%
\ref{ek}). Using this condition, the experimental value of $E_{0}=7$K \cite{schuster,allen} determines the size 
of the coherent dipole moment as: $\left| {\bf \mu }\right| \simeq e\cdot 0.01{\rm %
\AA }$. This value is indeed smaller than our previous estimation, which 
served as an upper bound on the size of the oscillating dipole moment (\ref{mu}). 
 
As we have proposed, the phase modulation in the (110) direction of the 
transverse atomic motion in the lattice, with energy $E(k)$ (\ref{ek}) 
should coincide with the T$_{1}$(110) phonon. In Fig.4 we compare the 
experimental values of T$_{1}$(110) taken from neutron scattering data with 
the calculated $E(k)$, and we find that the agreement is excellent for all $%
k $. From (\ref{ek}) and Fig.3 we see that at the edge of the Brillouin zone 
the energy $E(k)$ of the phonon should be just the bare energy of the local 
mode, $E_{0}$, since $X(\sqrt{2}\pi /a)=0$. We also have that at $k=\sqrt{2}%
\pi /a$ the dipoles have changed between the two configurations illustrated 
in Fig.3, which are the two possible configurations with alternating dipoles 
arranged on adjacent planes with the periodicity of the bcc unit cell. 
 
Since the empirical value of $E_{0}$ that we used was taken from NMR data, 
the agreement we find with the phonon data taken from inelastic neutron 
scattering, emphasizes the self-consistency of our description. We stress 
that the value of $E_{0}$ and the lattice vectors are the only empirical 
inputs used in the calculation, with the functional behavior completely 
given by the lattice structure and the dipolar interactions. 
 
\section{Off-Diagonal-Long-Range-Order and condensation.} 
 
We have found from the mean-field solution at zero temperature that the 
ground-state of the bcc phase contains a coherent-state of oscillating 
local-dipoles (\ref{psi0}). Since our method predicts the excitation 
spectrum of the T$_{1}$(110) phonon with very good accuracy, we expect it to 
be valid at the finite temperatures for which the bcc phase exists. We 
therefore expect that the basic nature of the bcc phase will be well 
described by our results, although the quantitative values may change due to 
the finite temperature. The coherent ground-state defines a global phase and 
breaks the gauge symmetry of a well-defined occupation number of local 
dipoles. In the limit $k\rightarrow 0$ we find that the occupation number of 
the local-modes diverges as $1/k$, signaling macroscopic Bose-Einstein 
condensation in the zero-momentum state  
\begin{equation} 
\left\langle n_{k}\right\rangle =v^{2}(k)=\frac{1}{2}\left( \frac{E_{0}{+X(k)%
}}{E(k)}-1\right) \rightarrow _{k\rightarrow 0}\frac{1}{2}\frac{E_{0}/2}{E(k)%
}=\frac{E_{0}/2}{2\hbar kc}  \label{nk} 
\end{equation} 
 
where $c$ is the sound velocity of the T$_{1}$(110) phonon which is the 
natural excitation of the dipolar array. This is identical to the result for 
a Weakly Interacting Bose Gas (WIBG) problem solved by Bogoliubov \cite{bogo}%
, where the divergence is related to the occupation of the zero-momentum 
state, i.e., the condensate fraction $n_{0}/n$
\begin{equation} 
\left\langle n_{k}\right\rangle _{WIBG}=v^{2}(k)\rightarrow _{k\rightarrow 0}%
\frac{n_{0}}{n}\frac{mc}{2\hbar k}  \label{nwibg} 
\end{equation} 
 
where in the WIBG case we have $c$ as the $k\rightarrow 0$ sound velocity, 
and $\varepsilon _{k}=\hbar ^{2}k^{2}/2m$ is the free particle energy. By 
comparing (\ref{nk}) with (\ref{nwibg}) we find that in the bcc case the 
role of the condensate-fraction, the WIBG order-parameter, is taken by the 
parameter $E_{0}$, which is just $2\left| X(0)\right| $. This can be seen 
directly from the form of the ground-state wavefunction (\ref{psi0}) where 
the pair-occupation is given by:  
\begin{equation} 
\left\langle b_{k}^{\dagger }b_{-k}^{\dagger }\right\rangle =2u(k)v(k)=\frac{%
X(k)}{E(k)}  \label{pair} 
\end{equation} 
 
Equating the divergent part in (\ref{nk}) and (\ref{nwibg}) we can define an 
effective condensate fraction 
\begin{equation} 
\frac{n_{0}}{n}=\frac{E_{0}/2}{mc^{2}}\simeq \frac{3.5}{10}=35\pm 8\% 
\label{nn0} 
\end{equation} 
 
where we used for the velocity of sound $c$ the values from our calculation (%
\ref{ek}) ($\sim 130m/\sec $) and from elastic constants \cite{minki} ($\sim 
160m/\sec $). It must be remembered that the mass $m$ in (\ref{nn0}) is not 
necessarily the mass of a bare $^{4}$He atom since we are now dealing with 
condensation of local dipoles. Comparing with the condensate fraction at 
zero temperature in superfluid $^{4}$He \cite{sokol}, which is $\sim 10\%$, 
we find that it is lower than the condensate fraction of the local modes in 
the bcc phase. We again mention that our result is for T=0 which can be 
depleted at finite temperature. 
 
It is clear that it is a non-zero coherent dipole moment $\mu $ that 
produces a dipolar interaction matrix $X(k)$ which in turn implies finite 
pair occupation (\ref{pair}) and a coherent ground-state. This is just the 
condition for the Bose-Einstein condensation of the dipoles in the bcc 
ground-state (\ref{nn0}). We therefore have a broken gauge symmetry and a 
complex order-parameter in the form of the pair-occupation (\ref{pair}). 
This function can be complex since the conditions on $u(k)$ and $v(k)$ allow 
for a relative complex phase between them, just as in the WIBG case. A 
similar condensation of local dipoles in all three orthogonal axes of local 
motion means that there are three independent phases at each lattice site, 
since orthogonal dipoles do not interact. The order-parameter in our case 
can therefore be described as a vector of three complex functions of 
independent magnitude and phase:  
\begin{equation} 
\Phi ({\bf r})=\left(  
\begin{array}{l} 
\left| \mu _{x}\right| e^{i\theta _{x}({\bf r})} \\  
\left| \mu _{y}\right| e^{i\theta _{y}({\bf r})} \\  
\left| \mu _{z}\right| e^{i\theta _{z}({\bf r})} 
\end{array} 
\right)  \label{order} 
\end{equation} 
 
If the cubic symmetry is not broken by external stresses, the magnitude
of the coherent dipole moment in the three orthogonal directions should
be the same: $\left| \mu _{x}\right|=\left| \mu _{y}\right|=\left| \mu _{z}\right|$.
In the ground-state the phases are spatially uniform, while the excited 
state is described through a periodic phase oscillation, i.e. the T$_{1}$%
(110) phonon. The order parameter (\ref{order}) is to be contrasted with the 
order parameter of superfluid $^{4}$He, which also exhibits ODLRO and which 
has a single complex component. 
 
In the hcp phase we do not expect the dipoles to order in a coherent state 
since the hexagonal geometry frustrates antiferroelectric-type 
configurations. Also the nearly isotropic potential of the hcp lattice
does not allow the highly directional dipole moments as in the bcc case. Indeed there is good agreement between experiments and the 
harmonic calculation of the phonons in the hcp phase \cite{minki2}, indicating no strong 
quantum corrections, as in the bcc phase. 
 
The bcc $^{4}$He is therefore a unique crystallographic phase having both 
Diagonal Long Range Order (DLRO) of the solid lattice and Off-Diagonal Long 
Range Order (ODLRO) of the local dipoles. It is not a 'super-solid' \cite 
{andreev,leggett,widom,stoof} in that it does not contain both a superfluid and 
a solid, but is more similar to the superconductors which have a DLRO of the 
atoms in the lattice and ODLRO\ of the superconducting electrons \cite{kohn}%
. This system is also distinct from the case of Bose-Einstein condensation 
of a phonon mode which results in a static deformation of the lattice and a 
structural phase-transition \cite{kohn}. 
 
Bose-Einstein condensation of local defects (vacancies) was previously 
considered for solid $^{4}$He \cite{andreev,leggett,widom}, and is similar 
to our treatment. The main difference is that in our case the physical 
picture of the condensed local modes is not a local distortion of the 
lattice like a vacancy, and that the condensation is unique to the bcc 
phase. The estimate in these works \cite{andreev,leggett,widom} is that in the 
ground-state (T=0) the density of vacancies is $\sim 10^{-4}$ per site (at 
molar volume of 21cm$^{3}$), and a condensate density of $\sim 10^{-7}$ per 
site. In contrast we expect in the bcc phase a sizable fraction (10-30\%) of 
condensed local-modes per site (\ref{nn0}). 
 
The fact that the region of existence of the proposed supersolid phase in 
the phase diagram of solid $^{4}$He should closely coincide with the region 
occupied by the bcc phase, was shown in \cite{stoof}. In this work it was 
further shown that in the supersolid there should be a second-sound-like 
mode, which is an oscillation in the density of the local-defects. In our 
description of the bcc phase this suggests the possibility of an oscillation 
in the amplitude of the order-parameter $\Phi ({\bf r})$ (\ref{order}), that 
is in the amplitude of the coherently oscillating dipole moment. This is in 
contrast to the T$_{1}$(110) phonon mode which is an oscillation in the 
phase of the order-parameter. This mode may be produced by modulating the 
density of the atoms so that the local excitation energies change and with 
them the amplitude of local motion and local electric dipole moment. 
Unfortunately we do not expect such a mode to have measurable consequences 
which are different from the effects produced by usual longitudinal phonons. 
 
In concluding this section, we would like to mention the recent experiments 
on the behaviour of implanted metalic ions (Cs) in solid $^{4}$He \cite 
{kanorsky}. These experiments are designed to look for evidence of 
time-reversal symmetry breaking which is equivalent to having a static 
electric dipole moment. In our description of the bcc phase we do not find a 
static but a coherent-dynamic electric dipole moment. We point out that in 
these experiments a marked difference between the hcp and bcc phases has 
been found. In the bcc phase the electronic-spin relaxation of the Cs atom 
is extremely slow and this effect could be a result of the coherence and 
long-range order of the dipolar fields. The coherently oscillating $^{4}$He 
electrons in the bcc phase will produce a very uniform electromagnetic 
interaction with the electronic spin in the Cs atom. By comparison, in the 
hcp phase the spin polarization is extremely short lived, indicating a more 
random field environment. This result is in accord with our expectation that 
the coherent dipoles are unique to the bcc phase. Similar experiments in the 
future may allow a probe that will show directly the coherently oscillating 
dipoles in the bcc ground-state. 
 
In these experiments \cite{kanorsky} the hyperfine transition in the Cs atom 
was also measured. The energy shift of this transition is sensitive to the shape 
of the confining cavity of the Cs atom inside $^{4}$He lattice. The width 
of the transition is a measure of the fluctuations in this cavity size \cite 
{kanorsky2}, and the data show a much smaller spread in the bcc compared 
with the hcp phase. Uncorrelated atomic motions of the $^{4}$He atoms will 
increase the spread in instantaneous cavity sizes due mainly to breating-like motion of the cavity walls (Fig.5).
This behavior is what we expect 
for the hcp case. The correlated atomic motion in the bcc phase should 
result in a more constant cavity shape (Fig.5) and a narrow signal, which is indeed measured \cite{kanorsky}. 
 
\section{Ground-state energy and the stability of the bcc phase} 
 
The question of the relative stability of the different crystal structures 
in solid He has been a long standing one. The necessity for some 
non-Van-der-Waals interactions has been previously proposed to explain the 
occurance of fcc over hcp structure in the heavier rare-gas solids \cite 
{venables}. The bcc phase is usually found to be more stable than the 
close-packed hcp phase due to the large zero-point energy in the He solids  
\cite{venables}. The correlations between the dipoles in the ground-state 
that we have proposed, lowers the energy of the ground-state of the bcc 
phase and further stabilizes it with respect to the hcp phase. The reduction 
in ground-state energy acheived by the coherent state of the dipoles along
one of the three orthogonal directions, is 
given by \cite{anderson}  
\begin{equation} 
\Delta E=\sum_{k}\frac{E(k)-(E_{0}+X(k))}{2E_{0}}<0  \label{dele0} 
\end{equation} 
 
which is negative since $X(k)<0$ and $E(k)<(E_{0}+X(k))$. 
 
Since the energy reduction integrand (\ref{dele0}) is non-zero only in the 
(110) directions it will give a small contribution to the three dimensional 
phase-space integration. At zero temperature the summation in (\ref{dele0}) 
will be confined to one dimensional sections along the (110) direction, so 
that the contribution will be zero. At the bcc temperatures ($\sim $1.4K) 
the one dimensional chains in the (110) directions are broadened so that the 
summation in (\ref{dele0}) is now over finite volume sections of phase 
space. We can estimate the maximum width of the conical section in $k$-space 
as the momentum which corresponds to a T$_{1}$(110) phonon with energy $%
k_{B}T$, that is $\sim 0.13$\AA $^{-1}$. The numerical integration of (\ref 
{dele0}) over such volume sections gives an energy reduction of $\Delta 
E\simeq -2$mK per atom. This result is in agreement with the experimentally 
interpolated energy difference between the bcc and hcp phases of solid $^{4}$%
He \cite{balibar}, which is of the order of a few mK per atom. 
 
This reduction is less than 0.1 percent of the potential and kinetic 
energies of the solid, and is therefore very hard to calculate accurately 
theoretically \cite{nosanow}. What is shown in the usual calculations is 
that the correlations between the motions of the atoms are essential in 
lowering the energy of the bcc phase, compared with the hcp phase. Since 
part of the correlations in the atomic motion is described by our coherent 
dipole model, we expect the condensation energy of the dipoles (\ref{dele0}) 
to be important in determining the stability of the bcc phase. At finite 
temperature the stabilization of the bcc phase compared to the hcp phase is 
usually attributed to the lower zero-point energy due to the lower T$_{1}$%
(110) phonon energy \cite{venables}. This is just the phonon which is 
softened by the long-range dipolar interactions that we have described, 
indicating again the importance of the coherent dipoles to the stabilization 
of the bcc phase of solid $^{4}$He. Our procedure may provide a good 
estimate of the small change in energy at the structural phase transition, 
by isolating the degree-of-freedom which is most affected by the transition, 
i.e. the correlated atomic motion along the directions normal to the unit 
cell faces. 
 
The picture we propose is that the dipole condensation mechanism of the bcc 
phase competes with the lower potential energy of the hcp phase due to its 
higher coordination number. If the hcp phase has a large enough volume 
(through thermal expansion or introduction of $^{3}$He impurities) its 
potential energy is increased until a critical point is reached where the 
bcc phase has a lower total energy due to the dipolar-condensation energy 
reduction (\ref{dele0}), which is absent in the hcp phase \cite{niremil}. At 
this critical point the structural phase transition occurs. 
 
By comparison, solid $^{3}$He has a stable bcc phase due to the larger 
kinetic energy of this lighter isotope. This increased zero-point energy 
causes the less dense bcc phase to have a lower ground-state energy than the 
hcp phase even at T=0. Since the $^{3}$He atoms are fermions with a spin 1/2 
nucleus, the oscillating electric dipoles are not in resonance and can not be 
treated as bosons \cite{glyde1}. We therefore do not expect a coherent state 
of the atomic motion as in the bcc $^{4}$He, and bcc $^{3}$He is stable due 
to its large zero-point kinetic energy alone. On the other hand, at low 
enough temperatures where the bcc $^{3}$He becomes an antiferromagnet, there 
could be correlations involving both the nuclear spin and electric dipole 
degrees of freedom. 
 
\section{Fermionic excitations} 
 
In addition to the fluctuations of the phase of the coherent dipole 
ground-state (i.e. T$_{1}$(110) phonons), there can be a localized 'flip' of 
a dipole so that it is in anti-phase (phase difference of $\pi $) relative 
to the rest of the dipoles, in the ground-state configuration. Such an 
excitation is naturally treated as a Fermion since such a flipped dipole is 
antisymmetric with respect to the other dipoles, that is with respect to the 
global phase $\theta $ (\ref{order}) in one (or more) of the orthogonal 
directions of local motion ($x,y,z$). The flipped dipole is no longer a 
dipolar image of the nearest-neighboring dipoles but an anti-image, and will 
be treated with Fermi-Dirac statistics. 
 
An anti-phase localized-mode (a fermion) is not part of the correlated 
ground-state, but nevertheless will feel the effect of the Bose excitations 
(T$_{1}$(110) phonons) of the dipolar array as they interact with it. The 
effective Hamiltonian describing such a fermion should therefore contain a 
term describing the creation and anihilation of pairs of fermions from the 
ground-state by a phonon (Boson). This is an off-diagonal term that 
describes the fluctuation caused by a T$_{1}$(110) phonon of energy $E(k)$: 
it changes a fermion 'particle' into a 'hole' and vice versa. The terms 
'particle' and 'hole' are with respect to the ground-state which has 
occupation of pairs of localized-modes (i.e. not an 'empty' vacuum). 
 
In addition there should be a term that describes the excitation energy of 
the bare fermionic localized-mode, that is $E_{0}$. This is just the energy 
to 'flip' a dipole from the ground-state so that all it's interactions with 
the other dipoles of the ground-state change sign, i.e. $-2X(0)=E_{0}$. 
 
The many-body effective Hamiltonian that we therefore propose is  
\begin{equation} 
H_{D}=\sum_{k}E(k)\left( c_{k}^{\dagger }c_{-k}^{\dagger 
}+c_{k}c_{-k}\right) -\sum_{k}V_{k}\left( c_{k}^{\dagger 
}c_{k}c_{-k}^{\dagger }c_{-k}\right)  \label{diracham} 
\end{equation} 
 
where $c_{k}^{\dagger },c_{k}$ are the creation and annihilation operators 
of the anti-phase (Fermionic) localized-mode. The first term in (\ref 
{diracham}) is the 'kinetic' term due to the phonon-roton branch, where the 
localized-modes are created/anihilated in pairs. The energy $E(k)$ is the 
energy of the T$_{1}$(110) phonon excitation (\ref{ek}). In addition there 
is a finite 'potential' energy if there is a finite density of unpaired 
fermions, which is $E_{0}$. In the absence of the second term we have just 
the Bose ground-state written in terms of fermionic pairs. 
 
We linearize the equations of motion that follow from (\ref{diracham}), 
similar to the BCS method \cite{kittel}  
\begin{equation} 
i\hbar \stackrel{\cdot }{c}_{k}=-E(k)c_{-k}^{\dagger }+\Lambda 
_{k}c_{k}\qquad i\hbar \stackrel{\cdot }{c}_{-k}^{\dagger 
}=-E(k)c_{k}-\Lambda _{k}^{*}c_{-k}^{\dagger }  \label{eqmotion} 
\end{equation} 
 
where we used the Fermi anti-commutation relations: $\left\{ 
c_k,c_k^{\dagger }\right\} =1\qquad \left\{ c_k,c_{-k}^{\dagger }\right\} =0$%
, and we define  
\begin{equation} 
\Lambda _k=\Lambda _k^{*}\equiv E_0\equiv \sum_kV_k\left\langle c_k^{\dagger 
}c_k\right\rangle ,\sum_k\left\langle c_k^{\dagger }c_k\right\rangle \equiv 
1,V_k=E_0  \label{linear} 
\end{equation} 
 
From (\ref{linear}) we see that the symbol $E_0$ will now indicate a finite 
density of fermions. 
 
The equations of motion have the following eigenvalues:  
\begin{equation} 
\left|  
\begin{array}{cc} 
E_{f}(k)-E_{0} & -E(k) \\  
-E(k) & E_{f}(k)+E_{0} 
\end{array} 
\right| =0\Rightarrow E_{f}(k)=\sqrt{E_{0}^{2}+E(k)^{2}}  \label{specdir} 
\end{equation} 
 
We can now solve the equations using the Bogoliubov-Valatin transformation 
for superconductivity \cite{kittel}:  
\begin{equation} 
c_{k}=u_{k}\alpha _{k}+v_{k}\alpha _{-k}^{\dagger }\qquad c_{-k}^{\dagger 
}=-v_{k}\alpha _{k}+u_{k}\alpha _{-k}^{\dagger }  \label{bogo} 
\end{equation} 
 
with the functions $u_{k},v_{k}$ given by  
\begin{equation} 
u_{k}^{2}=\frac{1}{2}\left( 1+\frac{E_{0}}{E_{f}(k)}\right) ,v_{k}^{2}=\frac{%
1}{2}\left( 1-\frac{E_{0}}{E_{f}(k)}\right)  \label{uvdir} 
\end{equation} 
 
The ground state is  
\begin{eqnarray}
\alpha _{k}\left| 0\right\rangle &=&0\Rightarrow \left| 0\right\rangle
=\prod_{k}\alpha _{-k}\alpha _{k}\left| vac\right\rangle \nonumber \\
&=&\prod_{k}\left(
u_{k}+v_{k}c_{k}^{\dagger }c_{-k}^{\dagger }\right) \left|
vac\right\rangle
\label{diracgs}
\end{eqnarray}
 
In Fig.6 we plot the energy spectrum (\ref{specdir}) compared with the other 
phonon modes in the (110) direction \cite{minki}. It is clear that this 
optic-like branch should be detectable in the low momentum range where it is 
not masked by the signal from the accoustic phonon modes. There is at 
present no high resolution neutron-scattering data in this energy and 
momentum range, and this prediction can be hopefully checked in future 
experiments. This mode could also be observed by Raman scattering, as a peak 
at energy $E_{0}$. 
 
\section{Scattering intensity} 
 
A\ neutron scattering inelastically from the solid will create/anihilate an 
elementary excitation. An excitation from the effective ground-states of the 
T$_{1}$(110) phonon (\ref{psi0}) and of the fermionic mode (\ref{diracgs}) 
involves an anihilation of a pair of local-modes, leaving an unpaired 
local-mode. We therefore expect the experimentaly measured neutron 
scattering intensity to be proportional to the density of local-mode 
pair-occupation at each wavevector $k$. This gives us the following results 
for the two modes: 
 
T$_1$(110) phonon:  
\begin{equation} 
I\propto \left\langle {{b_k}^{\dagger }}{b_{-k}^{\dagger }}\right\rangle =%
\frac{E_0}{E(k)}\left| \left( \frac{E(k)}{{E_0}}\right) ^2-1\right| 
\label{iphon} 
\end{equation} 
 
Fermionic excitation:  
\begin{equation} 
I\propto \left\langle c_{k}^{\dagger }c_{-k}^{\dagger }\right\rangle =\frac{1%
}{2}\frac{E(k)}{E_{f}(k)}=\frac{1}{2}\frac{1}{\sqrt{\left( E_{0}/E(k)\right) 
^{2}+1}}  \label{idirac} 
\end{equation} 
 
Both functions (\ref{iphon}),(\ref{idirac}) are plotted with arbitrary scale 
in Fig.7. We see that the intensity of the T$_{1}$(110) phonon is such that 
at small $k$ it behaves as $1/k$ which is typical for phonons at low $k$ and 
was seen experimentally \cite{minki2}, but goes identically to zero at the 
edge of the Brillion zone where $E(k)\rightarrow E_{0}$. The fermionic 
excitation has an opposite behavior by increasing linearly in intensity with  
$k$, until it saturates at the edge of the Brillion zone. The expression for 
the intensity of the T$_{1}$(110) phonon is similar to the expression of 
the intensity of the phonon-roton excitation spectrum of superfluid $^{4}$He  
\cite{nireric}, where it agrees very well with the experimental results. 
 
These predictions for the bcc phase have yet to be checked experimentally. 
 
\section{Conclusion} 
 
In this work we have investigated the nature of the quantum correlations in 
the bcc phase of solid $^{4}$He. We identified a three component complex 
order parameter and Bose-Einstein condensation in this phase, though not a 
'super-solid' \cite{kohn}, i.e. no superfluid component. There can be 
further manifestations of the ODLRO of the dipoles in the bcc phase which we 
have not explored yet, such as macroscopic topological defects in the 
complex order-parameter. The order-parameter or condensate-fraction can also 
serve as an extra thermodynamic variable, and this opens the possibilty of 
more complicated internal dynamics in the bcc solid, such as the phenomenon 
of second sound in superfluid $^{4}$He. 
 
We predict that a local excitation of a dipole out of the coherent 
ground-state will behave as a Fermion, and we calculate its energy spectrum. 
We find it to behave as an optical-like branch in the (110) direction. 
Finally we calculate the scattering intensity as a function of wavevector $k$ 
for both the Bose (T$_{1}(110)$ phonon) and Fermi (new optical-like branch) 
excitations. All these predictions await high-resolution neutron and Raman 
scattering experiments to be compared with. 
 
{\bf Acknowledgements} 
 
I thank Emil Polturak for useful discussions and encouragement. 
 
This work was supported by the Israel Science Foundation and by the Technion 
VPR fund for the Promotion of Research. 

\newpage
 
\section{Appendix A: Comparison of Bose excitations with Klein-Gordon Hamiltonian.}

We would like to point out that the Hamiltonian describing
the localized dipoles (%
\ref{hloc}) is similar to the Klein-Gordon (KG) Hamiltonian
for a
single spinless boson, written in its first-order form \cite{baymq}:

Klein-Gordon:
\begin{equation}
H_{KG}=\varepsilon _k\left( \sigma _z+i\sigma _y\right) +mc^2\sigma
_z+e\Phi
\widehat{1}  \label{hamkg}
\end{equation}

Localized dipoles (\ref{hloc}):
\begin{equation}
H_{loc}=X(k)\left( \sigma _z+i\sigma _y\right) +E_0\sigma _z
\label{hamdip}
\end{equation}

where $\sigma _i$ are the Pauli matrices, $\varepsilon
_k=\widehat{p}^2/2m$, $%
e$ is the electric charge, $\Phi $ is the electrostatic potential, $m$
is
the KG-particle's mass and $c$ is the velocity of light. 

We have written the dipolar hamiltonian (\ref{hamdip}) in the basis of a two
component wavefunction 
\begin{equation}
\Psi _{loc}=\left(
\begin{array}{c}
c_k^{\dagger } \\
c_{-k}
\end{array}
\right)
\label{dipolewave}
\end{equation}

In this representation we see that exciting a local dipole out of the
ground-state configuration ($c_k^{\dagger }$) has bare energy $E_0$ while
destroying an excited dipole has minus this energy.
There is a freedom of choice weather to define the positive excitation to be a
flipping of an up dipole to a down dipole or vice versa. The sign of the energy of
the dipolar bosons therefore represents this freedom which corresponds to two
equivalent dipolar configurations with a $\pi$ phase difference.

The
two-component
wavefunction of the KG Hamiltonian is:
\begin{eqnarray}
\Psi _{KG}&=&\left(
\begin{array}{c}
\varphi \\
\chi
\end{array}
\right) \nonumber \\
\varphi &=&\frac 12\left( \psi +\frac{i\hbar }{mc^2}\psi
^0\right)
,\chi =\frac 12\left( \psi -\frac{i\hbar }{mc^2}\psi ^0\right)
\label{kleinwave}
\end{eqnarray}

where $\psi $ is the original wavefunction of the second-order KG
equation,
and $\psi ^0=\left( \frac \partial {\partial t}+\frac{ie}\hbar \Phi
\right)
\psi $. 
The Hamiltonians (\ref{hamkg},\ref{hamdip}) are the similar 
except that the KG density is not normalized to 1 but to $\left\langle
\rho
\right\rangle =E/mc^2$, describing the relativistic increase in the
density
with velocity.
By comparing the two Hamiltonians (\ref{hamdip},\ref{hamkg}) we identify that: 
$E_0\leftrightarrow mc^2$, $X(k)\leftrightarrow \varepsilon _k$,
which gives the equivalence of the two hamiltonians.

The peculiarities of the KG equation appear when there is a potential $%
V=e\Phi $ (the Klein paradox for example). The equation for the momentum
of
the KG prticle:
\begin{eqnarray}
2mc^2\varepsilon _k&=&\hbar ^2c^2k^2=\left( E(k)-V\right) ^2-\left(
mc^2\right)^2  \nonumber \\
\Rightarrow k&=&\frac{\sqrt{\left( E(k)-V\right) ^2-\left( mc^2\right)
^2}}{\hbar c}  \label{kgk}
\end{eqnarray}

becomes the equation for $X(k)$ in the dipolar case:
\begin{equation}
X(k)=\frac{\left( E(k)-V\right) ^2-\left( E(k) _0\right) ^2}{2E_0}
\label{xkv}
\end{equation}

We see from (\ref{xkv}) that there is a region of energies where the
interaction parameter $X(k)$ is positive and a region where it is
negative.
We saw above that the condensation of the dipoles in the bcc phase is characterized by a negative $X(k)$
which
also gives a gapless excitation
spectrum
at $k\rightarrow 0$. The excitations with $%
E(k)>V+E_0,E(k)<V-E_0$ are therefore not contributing
to
the coherent long-range order.
A fermionic excitation is a local destruction of the coherent order
, and indeed costs at least $E_0$ (for the free case with $V=0$)
to
create (\ref{specdir}).

In the case of the KG equation the sign of the enrgy indicates the charge of
the particle/antiparticle, which have oposite charges.
Charge conjugation therefore interchanges between the two.
What is the meaning of the different signs of the energy
of the dipolar bose excitations in the bcc case ? From our definition of the second-quantized
description of the dipoles, the meaning of the sign of the energy is that the field of
resonating localized-dipoles can have two global configurations shifted
by $%
\pi $ (Fig.2). These two configurations are identical with respect to the magnitude
of the energy
spectrum, but in each the operation of spin flip changes from
up$\rightarrow$down to down$\rightarrow$up. We can therefore identify two ''charges'' for the bcc
to
distinguish between the two shifted phases. Further we find that as in the KG
case the operation of charge conjugation (which reverses the signs of the
dipoles) moves us between the two solutions. 

\section{Appendix B: Symmetry breaking of the fermionic excitations.}

We see from (\ref{uvdir}) that when there is no fermion present (i.e. if
we
put $E_{0}=0$ in (\ref{linear})) the ground state has equal numbers of
fermions and holes. The symmetry between particles and holes is broken
by
the free fermion quasiparticle (or quasihole), and the sign of the symmetry-breaking
parameter $%
E_{0}$ determines which of the two kinds is present. The hole/particle
are
with respect to the equilibrium occupation by pairs of fermions in the
ground-state.

A single flipped dipole described as the fermionic excitation, breaks
the
symmetry between the number of up/down dipoles and creates a residual globally 
oscillating dipole moment. The parity $P$ symmetry with respect to
reflection along the axis of the global dipole (let us choose to be $z$)
is
broken. The charge $C$ symmetry is also broken since the direction of the global dipole
is
flipped under charge conjugation. The time reversal symmetry $T$ is
unbroken
since the oscillating globel dipole does not define a unique time
direction.
We therefore have that the global $CPT$ symmetry is preserved, as is the
$CP$
and $T$ symmetries individually. The symmetry-breaking parameter in
(\ref
{diracham}) is the sign given to $E_{0}$, which corresponds to choosing
an
up or down dipole to flip. In second quantization langauge this is the
choice between an unpaired particle or hole. The broken symmetry is not
of
the $U(1)$ group such as the $\Phi $ (\ref{order}) order parameter, but
has
a $Z(1)$ discrete symmetry.

We now compare this with the situation of the two-dimensional massive
Dirac
particle \cite{berry}. The Hamiltonian describing a single fermionic
excitation (\ref{diracham}) can be written as
\begin{eqnarray}
i\hbar \frac{\partial }{\partial t}\left(
\begin{array}{c}
c_{k} \\
c_{-k}^{\dagger }
\end{array}
\right) &=&
E(k)\left(
\begin{array}{cc}
0 & 1 \\
1 & 0
\end{array}
\right) \left(
\begin{array}{c}
c_{k} \\
c_{-k}^{\dagger }
\end{array}
\right)  \nonumber \\
&+&E_{0}\left(
\begin{array}{cc}
1 & 0 \\
0 & -1
\end{array}
\right) \left(
\begin{array}{c}
c_{k} \\
c_{-k}^{\dagger }
\end{array}
\right)
\label{eqmotionmat}
\end{eqnarray}

while the 2D Dirac particle is described by
\begin{equation}
i\hbar \frac{\partial }{\partial t}\left(
\begin{array}{c}
\varphi \\
\chi
\end{array}
\right) =c(\widehat{{\bf \sigma }}\cdot {\bf p})\left(
\begin{array}{c}
\varphi \\
\chi
\end{array}
\right) +mc^{2}\widehat{\sigma }_{z}\left(
\begin{array}{c}
\varphi \\
\chi
\end{array}
\right)  \label{diracmat}
\end{equation}

By assuming momentum ${\bf p}$ in the $\widehat{x}$-direction only we
write
\begin{eqnarray}
i\hbar \frac{\partial }{\partial t}\left(
\begin{array}{c}
\varphi \\
\chi
\end{array}
\right) &=&c\left(
\begin{array}{cc}
0 & -i\hbar \partial _{x} \\
-i\hbar \partial _{x} & 0
\end{array}
\right) \left(
\begin{array}{c}
\varphi \\
\chi
\end{array}
\right) \nonumber \\
&+&mc^{2}\left(
\begin{array}{cc}
1 & 0 \\
0 & -1
\end{array}
\right) \left(
\begin{array}{c}
\varphi \\
\chi
\end{array}
\right)
\label{diracmat2d}
\end{eqnarray}

where $\widehat{{\bf \sigma }}=\left( \widehat{\sigma
}_{x},\widehat{\sigma }%
_{y},\widehat{\sigma }_{z}\right) $, $\widehat{\sigma }_{i}$ the Pauli
matrices, and $\varphi ,\chi $ are the particle/antiparticle scalar
wavefunctions. There is now complete analogy between
(\ref{eqmotionmat})
and (\ref{diracmat2d}). The symmetry-breaking parameter $E_{0}$ is
identical
to the $mc^{2}$ parameter in the 2D Dirac equations-of-motion (\ref
{diracmat2d}). The symmetry that is broken by choosing a non-zero
$mc^{2}$
in two-dimensions is the time-reversal symmetry (TRS) \cite{berry}. As
shown
in (\ref{diracmat}), the time component of the momentum-energy vector in
two
dimensions is taken by the $z$ axis. The broken parity $P$ in the
$z$-axis
for the fermionic excitation of the bcc phase is here replaced by the
TRS
breaking of the heavy two-dimensional Dirac particle.

Similar to the Anderson \cite{andersonbcs} transformation of the BCS
problem
to a magnetic Hamiltonian we can transform (\ref{diracham}) using:
\begin{eqnarray}
n_k &=&c_k^{\dagger }c_k\qquad c_k^{\dagger }c_{-k}^{\dagger }=\sigma
_k^{-}/2\qquad c_kc_{-k}=\sigma _k^{+}/2  \nonumber \\
&\Rightarrow &n_kn_{-k}=\frac 12\left( \sigma _k^z+1\right) \qquad
c_k^{\dagger}c_{-k}^{\dagger }+c_kc_{-k}=\sigma _k^x
\label{magtrans}
\end{eqnarray}

where the $\sigma _k^i$ are Pauli spin-1/2 operators. The basis is such
that
an up-spin in the $\widehat{z}$-direction represents an empty pair,
while a
down-spin represents an occupied pair.

The resulting Hamiltonian is:
\begin{equation}
H_{mag}=\sum \varepsilon _k\sigma _k^x-\frac 12\sum V_k\left( \sigma
_k^z+1\right)  \label{hamag}
\end{equation}

This Hamiltonian describes a fictitious magnetic field acting on the
spin $%
\overrightarrow{\sigma }$:
\begin{equation}
\overrightarrow{B}=\varepsilon _k\widehat{x}-\frac 12E_0\widehat{z}
\label{mag}
\end{equation}

where we replaced the potential energy with the constant $V_k=E_0$
(\ref{linear}).
The magnetic field (\ref{mag}) can be compared with the BCS result
\cite{andersonbcs}
\begin{equation}
\overrightarrow{B}_{BCS}=\varepsilon _k\widehat{z}+\frac 12V\sum \left(
\sigma _k^x\widehat{x}+\sigma _k^y\widehat{y}\right)  \label{magbcs}
\end{equation}

The alignment of the spins in the ground-state is shown for the two
Hamiltonians in
Fig.8.
In the BCS problem the sign of the symmetry-breaking field in $V$ (\ref
{magbcs}) has to be positive so that it induces ferro-magnetic interaction
between the fictitious spins, otherwise there will not be any rotation of the spins
across the Fermi-energy. Only when the spins rotate do they go through
the
point where the spin is entirely in the $xy$-plane. At this point the
state
has no defined occupation number but a well defined phase, while on both
sides of the domain-wall there is well defined occupation and no phase.
This
superconducting-phase at the fermi energy is just the angle of the fictitious spin in the $xy$-plane,
and
there is broken U(1) symmetry.

In the Dirac case the symmetry-breaking field $E_{0}$ in (\ref{mag}) is
a
constant external field. It can have both signs, which control the
direction
along the $z$-axis that the rotated spin has in the middle of the
domain-wall. This spin describes weather a particle or a hole is
occupied,
while away from the $k=0$ point the spins are in the $xy$-plane, with a
well
defined phase. The symmetry that is broken is therefore the binary Z(1) ($\pm
$) symmetry.

In comparison with the BCS problem we see that in the fermionic excitation in
the bcc the
symmetry-breaking parameter is a finite density of unpaired fermions: $%
\left\langle c_{k}^{\dagger }c_{k}\right\rangle \neq 0$ (\ref{linear}). The
ground-state
without unpaired fermions is a 'vacuum' of pairs of particle-holes in
equal
numbers. In the BCS problem the symmetry-breaking parameter is a finite
pair-density: $\left\langle c_{k}^{\dagger }c_{-k}^{\dagger
}\right\rangle
\neq 0$. The ground-state in the absence of electron pairing is just a
finite density of electrons below the fermi-energy and zero above. In
this
respect the two problems are 'complementary'.

\begin{figure}[tbp]
\input epsf \centerline{\ \epsfysize 11.5cm \epsfbox{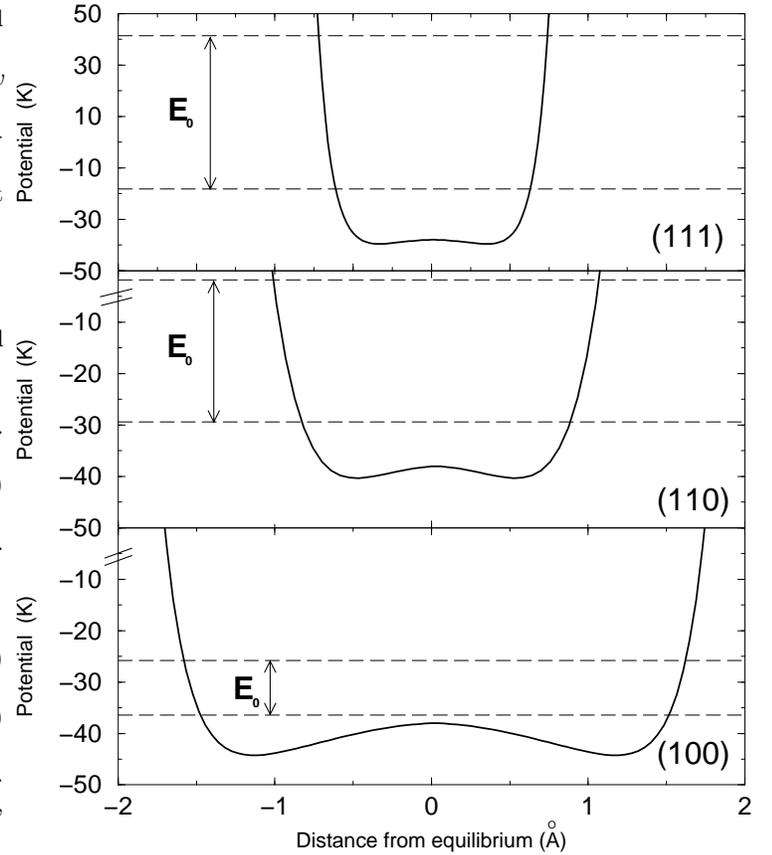}}
\caption{The potential-well of an atom in bcc $^{4}$He along different
directions. The energy difference $E_{0}$ between the lowest two energy
levels (dashed lines) are: (111)- $59.5$K, (110)- $27.6$K, (100)- $%
10.6$K.}
\end{figure}
\begin{figure}[tbp]
\input epsf \centerline{\ \epsfysize 5.5cm \epsfbox{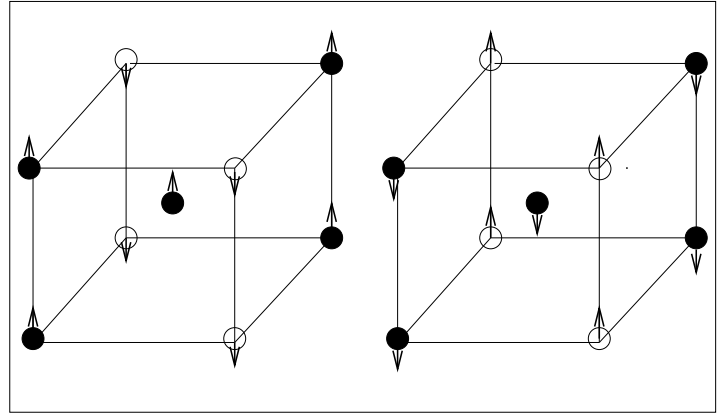}}
\vskip 3mm
\caption{The coherent dipole arrangement in the ground-state of the
bcc
phase, oscillating between these two configurations. Dipoles with same
phase
have same shade. The sum of the dipole-dipole interaction (Eq.3) for a
unit dipole-moment is: -0.08 (${\rm \AA }^{-3}$).}
\end{figure}
\newpage
\begin{figure}[tbp]
\input epsf \centerline{\ \epsfysize 10.5cm \epsfbox{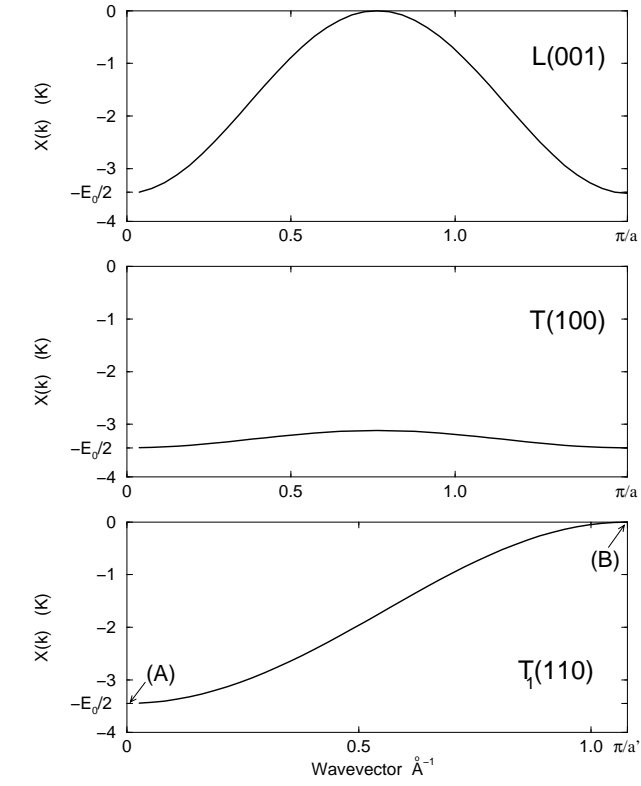}}
\vskip 3mm
\input epsf \centerline{\ \epsfysize 4.0cm \epsfbox{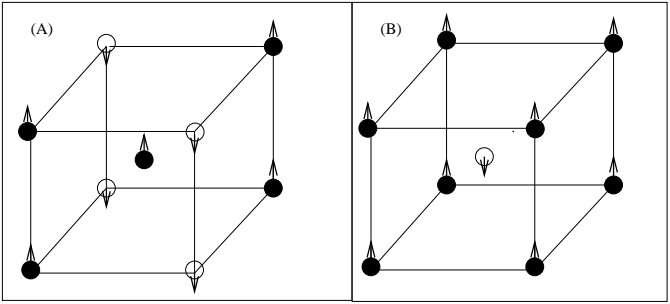}}
\vskip 3mm
\caption{The calculated interaction matrix $X(k)$ (Eq.4) as a function
of the
wavevector $k$, for the three phonon modes that could affect a dipolar
array. The dipole moment has been normalized to give a gapless mode:
$X(k=0)=-E_{0}/2$.
The unit-cell dimensions $({\rm \AA })$: $a=4.12/2$, $a'=a
\protect{\sqrt{2}}$.
Also shown are the two arrangements of the dipoles in the extreme points
along the oscillation in the (110) direction.}
\end{figure}
\begin{figure}[tbp]
\input epsf \centerline{\ \epsfysize 10.5cm \epsfbox{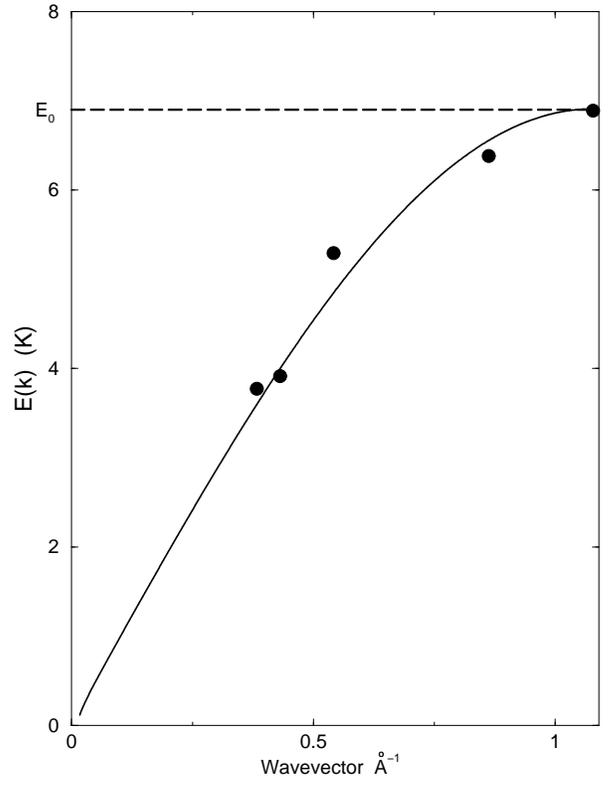}}
\caption{The experimental data [13] (solid circles) for the T$_{1}$(110)
phonon compared with the calculation (Eq.8) (solid line). Also shown is
the
energy of the bare local-mode $E_{0}=7$K.}
\end{figure}
\begin{figure}[tbp]
\input epsf \centerline{\ \epsfysize 4.5cm \epsfbox{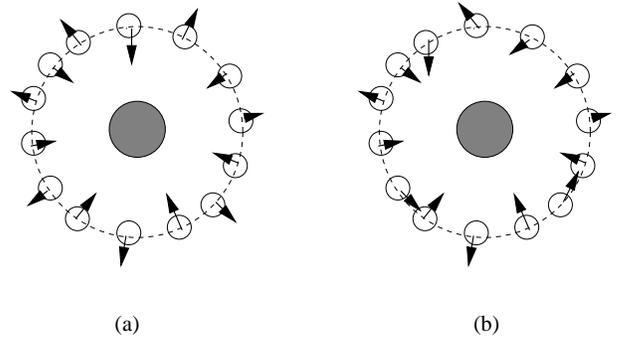}}
\vskip 3mm
\caption{Schematic cavity shapes for a Cs atom (grey circle) inside $^4$He solid (empty circles with arrows). (a)
Correlated atomic motion in bcc: constant cavity shape, (b) uncorrelated
atomic motion in hcp: randomly fluctuating cavity shape.}
\end{figure}
\begin{figure}[tbp]
\input epsf \centerline{\ \epsfysize 7.0cm \epsfbox{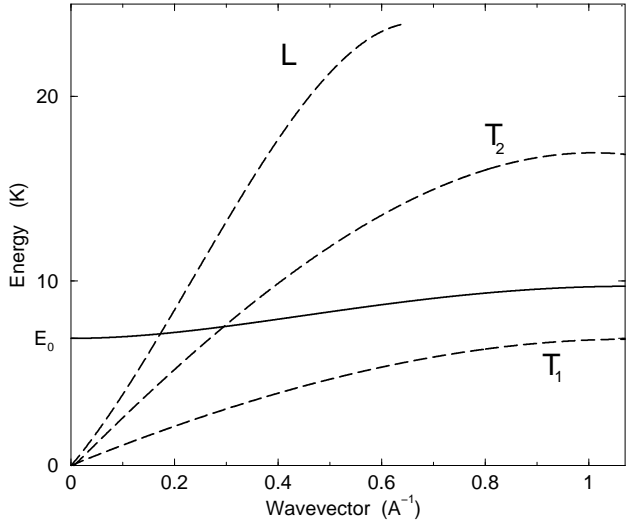}}
\caption{The spectrum of the fermionic optic-like mode (Eq.18, solid
line) compared with the experimentally measured phonons in the (110)
direction [13].}
\end{figure}
\begin{figure}[tbp]
\input epsf \centerline{\ \epsfysize 10.5cm \epsfbox{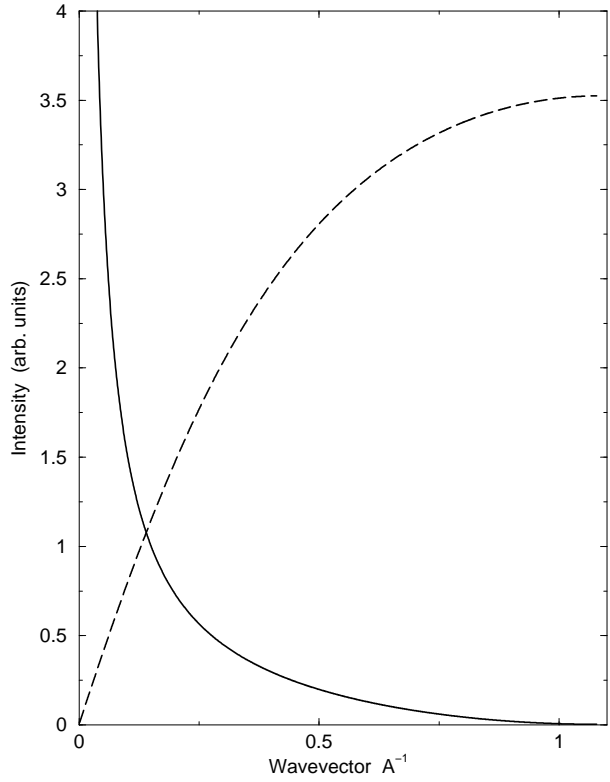}}
\caption{The calculated intensity of the inelastic neutron scattering by
the
T$_{1}$(110) phonon (Eq.22, solid line) and by the new fermionic mode
(Eq.23, dashed line).}
\end{figure}
\begin{figure}[tbp]
\input epsf \centerline{\ \epsfysize 9.5cm \epsfbox{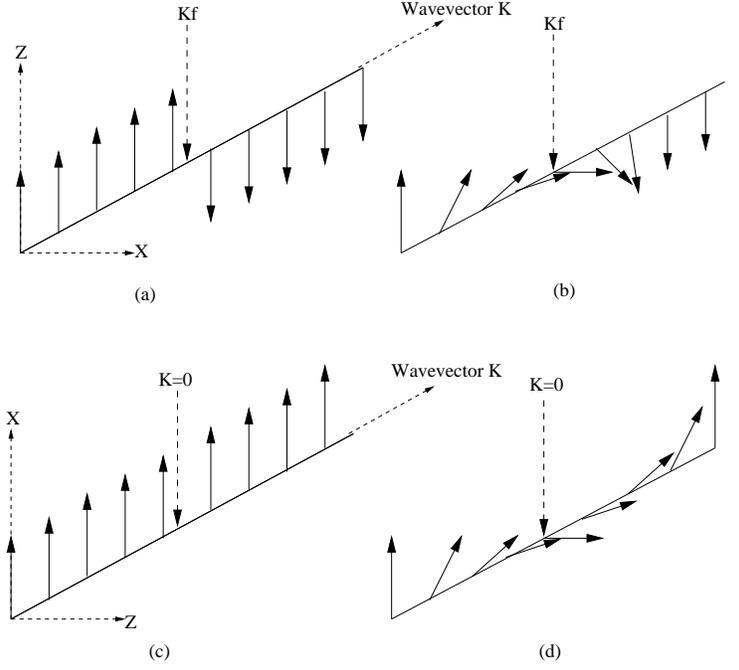}}
\vskip 3mm
\caption{Arrangement of the spins in the magnetic-analogue-Hamiltonian.
BCS: (a) without pairing, (b) with pairing. Dirac: (c) without unpaired
fermion, (d) with unpaired fermion.}
\end{figure}

\end{document}